\documentclass{article}
\usepackage{frascatiphys,here,graphicx,subfigure}
\begin{document}
\title{
THE MANY USES OF CHIRAL EFFECTIVE THEORIES
}
\author{
Elisabetta Pallante \\
{\em Centre for Theoretical Physics, University of Groningen, The Netherlands\footnote{email: e.pallante@rug.nl}}
}
\maketitle
\baselineskip=11.6pt
\begin{abstract}

I review basic concepts of chiral effective field theories guided by an historical 
perspective: from the first ideas to the merging with other effective frameworks, and to 
the interplay with lattice field theory. The impact of recent theoretical developments on 
phenomenological predictions is reviewed with attention for rare decays, and 
charm physics. I conclude with a critical look at future applications.

\end{abstract}
\baselineskip=14pt
\section{ A retrospective}
Effective field theories are the protagonists of our modern view of quantum field theory. 
The idea that any sensible theory is a priori valid only on a limited interval of energies, 
else distances, became more and more accepted during the last decades. Any such theory 
carries a dependence on a particular high energy scale, {\em the ultraviolet 
cutoff} which determines its range of validity. It contains the low energy or large 
distance behaviour of a more fundamental theory. Only the ultimate fundamental theory, 
if any, must be valid at all energies or distances.
This broadened view led us to abandon the concept of renormalizability in a strict sense as 
the necessary requisite for a theory to be an acceptable theory. 
Effective field theories are by now one of the main 
theoretical instruments for exploring a large set of particle physics phenomena, from the 
very low-energy strong interactions to the candidate models for the physics beyond the 
electroweak symmetry breaking scale.
In this writeup I will care for concepts more than numbers, and make use of what an 
historical perspective can teach us.

Effective field theories (EFT) started as {\em phenomenological lagrangians}, aimed at 
describing the dynamics of strongly interacting matter, mesons and baryons, at low energy. 
In general, they were aimed at describing any system where the dynamics is governed by a 
given internal symmetry and its spontaneous breaking. The original works date in the 1960's, 
mainly by Schwinger, Cronin, Weinberg\cite{Cronin,WPRL,WPR}, and the works by Callan, Coleman, Wess, Zumino\cite{CWZ,CCWZ}.
The structure of phenomenological lagrangians was purely based on symmetries and much 
inherited from current algebra; here, the low energy strong interactions could be described 
by a formulation alternative to Quantum Chromodynamics (QCD)\footnote{Phenomenological lagrangians evolved together with the concept of {\em quarks} degrees of 
freedom\cite{GellMannZweig} and the description of strong interactions with a non abelian 
gauge theory}, the theory with quarks and gluons degrees of freedom.

Chiral Perturbation theory (ChPT) was formulated more than a decade later by Gasser and 
Leutwyler in two by now well known papers\cite{GL1,GL2} in 1984 and 1985.
ChPT is the descendant of phenomenological lagrangians. It is a particular example of a 
non-decoupling effective theory. Its fundamental symmetry 
is the chiral symmetry, with group $SU(2)_L\times SU(2)_R$ or $SU(3)_L\times SU(3)_R$\footnote{I will omit the subscripts $L,R$ in the following.}, spontaneaously broken 
down to its diagonal subgroup. The derivation of the ChPT lagrangian and properties by 
a path integral formulation\cite{GL1} clarifies its field theoretical connection to QCD. 
Somewhat more recently Heavy Quark effective theory (HQET) 
was introduced as a good theoretical 
approximation to describe the dynamics of systems with one heavy quark\cite{HQET}.
It is the merging of these two formulations that gave rise to new types of effective field
theories, namely the Heavy Baryon ChPT\cite{HBChPT} (HBChPT) for describing the interactions amongst 
baryons and light mesons, and the Heavy-Light Meson ChPT\cite{HLChPT} (HLChPT) for 
describing the 
dynamics of bound meson systems such as $D$, $D_s$, $B$, and $B_s$.

\begin{figure}[H]
 \begin{center}
 {\includegraphics[scale=0.29]{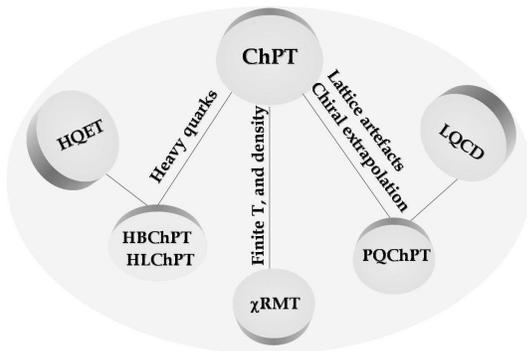}}
 \caption{\it The merging of ChPT with HQET and lattice QCD gives rise to new effective
 field theoretical descriptions of physical phenomena like baryon and heavy-light meson
 physics, 
field theory at finite volume, on discretized spacetime and in the (partially) quenched 
limit (PQChPT). Another branch in the figure points at the particularly interesting 
realization of Chiral Random Matrix Theory\cite{cRMT} ($\chi$RMT) recently proved to be 
equivalent to ChPT\cite{BHO}, a powerful tool to explore 
QCD at finite temperature and nonzero chemical potentials.}
\label{figure1}
 \end{center}
\end{figure}
Moving forward in time we encounter a fruitful interplay of this wide class
of  EFT descriptions with the lattice formulation of field theories and in particular of 
strong interactions (referred to as Lattice QCD (LQCD)). During the last years it has 
become clear how the complementary use of both 
approaches is extremely useful to understand non perturbative aspects of a field 
theory, possibly gaining insight into the way for an exact solution; one exciting example is 
the attempt at extending the AdS/CFT conjecture\cite{Maldacena} to AdS/QCD\cite{Shuryak}.  

In the following sections I review the basic principles of phenomenological 
lagrangians, their descendants, and discuss a few topics 
in the phenomenolgy of hadron interactions, where the role of EFT and LQCD is and will be 
especially relevant. I conclude with some thoughts on possible future developments and 
applications.  

\section{The Theory}
The formalism of phenomenological lagrangians was mainly motivated by the necessity of 
describing the interactions of phenomenological fields, like the pions, whose appearance 
is due to the spontaneous breaking of an internal global symmetry. 
The mathematical problem is equivalent to that of finding all nonlinear realizations of a 
(compact, connected, semisimple) Lie group which become linear when restricted to a given 
subgroup\cite{CWZ}. The following problem is the one of constructing nonlinear 
lagrangian densities which are invariant under the nonlinear field 
transformations\cite{CCWZ}.

Consider the chiral group $G=SU(N)\times SU(N)$ which is spontaneously broken down to the 
diagonal (parity-conserving) subgroup $H=SU(N)_V$. The pattern of symmetry breaking is 
$SU(N)_L\times SU(N)_R \to SU(N)_V$ with $N=2,3$ flavours.
Of the total number of generators of G, there will be $N^2-1$ ``exact'' generators of the 
subgroup H, and $N^2-1$ ``broken'' 
generators of the residual subgroup. Fields are associated to the generators of $G$, and 
pions, the Goldstone bosons of the spontaneaously broken chiral symmetry,
 are associated to its broken generators.
Current algebra was implying that to correctly describe pion interactions it was necessary 
to eliminate all non derivative couplings from the lagrangian. The mathematical solution 
and the construction of the correct lagrangian for pions
was reached in two ways: from the old $\sigma$-model, by performing\cite{WPRL} a chiral 
rotation of the four-dimensional field $(\sigma ,{\bf\pi})$ of $SU(2)$, which eliminates the 
non derivative
coupling of $\sigma$ and ${\bf{\pi}}$ and replaces it with a nonlinear derivative 
coupling of the chiral rotation vector, identified as the new pion field and transforming 
as a nonlinear realization of $G$.\footnote{The fact 
that one can limit the field content to a pion triplet and does not need to add a scalar 
field is due to the existence of a three-dimensional non linear realization of $SU(2)\times
 SU(2)$ while there is no three-dimensional linear representation.}
The second more elegant way\cite{WPR,CWZ,CCWZ} was to directly postulate the nonlinear 
transformation properties of the pion fields and to construct a $G$ (chiral) invariant 
lagrangian.
The recipe\cite{CCWZ} for such a lagrangian amounts to
\begin{equation}
{\cal L} = c Tr\, \left [\partial_\mu \left ( e^{i\pi\cdot T}\right )\partial^\mu \left 
( e^{-i\pi\cdot T}\right )\right ]\, ,
\end{equation}
where $\pi\cdot T = \sum_{i=1}^{N^2-1}\, \pi_i\, T_i$, with $T_i$ the broken generators and 
$\pi_i$ the associated pion fields. The coupling constant $c$ is proportional to the scale
 of the symmetry breaking.
This lagrangian describes self-interactions of the phenomenological fields $\pi$, and it 
is nonlinear in the fields: its exponential dependence generates infinitely many interaction 
terms. Another peculiarity is that it contains only derivative type interactions, which 
means that at low energies the fields are weakly interacting.   
In a paper of 1979\cite{Weinberg} entitled ``Phenomenological Lagrangians'', Weinberg 
constructed the renormalization group relations amongst the divergent structures appearing 
in the loop expansion of the theory. These relations clarify in which sense the theory 
is nonrenormalizable. 

The formulation of ChPT appeared in two seminal papers by Gasser and Leutwyler; in the 
first\cite{GL1} the $SU(2)$ flavour theory is derived, and in the second\cite{GL2} the 
theory is extended to SU(3) to include the heavier strange quark. 
The lagrangian 
\begin{eqnarray}
{\cal L}_2 &=&  \frac{f_\pi^2}{4}\, Tr\, (D_\mu \Sigma D^\mu \Sigma^\dagger )\,\, +\,\,  
\frac{f_\pi^2}{4}\, Tr\, (\chi^\dagger \Sigma +\chi\Sigma^\dagger )
\nonumber
\end{eqnarray}
is the first order contribution to an expansion in powers of the small energies of the 
fields $\Sigma = e^{\frac{2i}{f_\pi}\, \Phi\cdot T}$ and the light quark masses, which are
 invariantly introduced through the scalar spurion field $\chi = 2B_0 {\cal M} + \ldots $, 
with 
$M=diag(m_u,m_d,m_s)$ and $B_0$ the parameter proportional to the scalar quark condensate. 
Covariant derivatives are defined to contain external vector and axial spurion 
fields $D_\mu\Sigma =  \partial_\mu \Sigma -i(v_\mu +a_\mu)\Sigma +i\Sigma (v_\mu - a_\mu)$.

ChPT is the effective description of a strongly coupled theory, which is low energy QCD.
Its expansion in powers of small momenta and light quark masses
\begin{eqnarray}
{\cal L} &=& {\cal L}_2\,\, +\,\, \frac{1}{{\Lambda_\chi}^2}{\cal L}_4\,\,+\,\, 
\ldots~~~~~~~~~~~~ p^2\sim M_\pi^2\sim m_q
 \nonumber
\end{eqnarray}
has a predictive power which is dictated by the numerical value of its ultraviolet cutoff, 
by construction the scale of spontaneous chiral symmetry breaking $\Lambda_\chi \propto
 f_\pi$, the pion decay constant. The equality $\Lambda_\chi \simeq 4\pi 
f_\pi \simeq 1\, GeV$\footnote{ It is suggested by the numerical behaviour of the loop 
expansion, and not derived from first principle considerations.} guarantees a good
 predictivity for energies well below 1 GeV. 

Flavour physics involving dominant contributions from low energy strong interactions can 
be explored with ChPT: the SU(2) case completely describes pion physics and the 
physics of isospin breaking for $m_u\neq m_d$. The SU(3) case describes kaon physics, 
with $m_u,m_d\ll m_s$ and provided the kaon mass $M_K\ll \Lambda_\chi$. 
ChPT plays an essential role in the calculation of QCD induced corrections to weak decays of 
light mesons. Golden channels are certainly the nonleptonic kaon decays, source of
the $\Delta I =1/2$ rule and probe of indirect CP violation through 
$\varepsilon'/\varepsilon$, and rare kaon decays, useful to constrain sources of new 
physics beyond the standard model. There are special cases where large corrections to 
$SU(3)$ processes are purely $SU(2)$ effects, as it is for final-state-interactions in 
$K\to\pi\pi$ decays\cite{PP}.
\begin{figure}[H]
 \begin{center}
 {\includegraphics[scale=0.33]{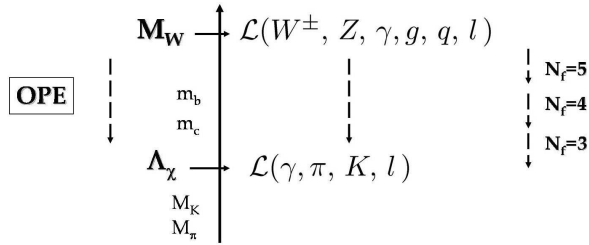}}
 \caption{\it Flow diagram of the unified EFT description of weak and strong interactions. 
$N_f$ is the number of active flavours in a given theory. By decreasing energies, the top 
quark and $W$ boson, the bottom quark, the charm quark are subsequently ``integrated out''. 
$\Lambda_\chi$ is the chiral symmetry breaking scale at which the nonperturbative matching 
of ChPT with QCD is performed.}
\label{Figure2}
 \end{center}
\end{figure}
The unified description of weak and strong interactions as effective field theories is the 
perfect example of the two ways in which an EFT can be realized. Weak interactions identify 
with a weakly coupled theory, where decoupling (of massive modes) takes place and 
perturbative matching can be performed. In a sequence of decreasing energies 
(see fig.\ref{Figure2}), starting at
 the mass of the $W$ boson, the renormalization group equations and Operator Product 
expansion (OPE) run the theory to lower energies. Massive modes ``decouple'' from the theory 
at each threshold\footnote{The way decoupling manifests depends on the renormalization 
scheme used. Smooth decoupling does not arise in the typically used $\overline{MS}$ scheme 
where the decoupling consists of ``integrating out'' the corresponding massive particle.}
and give rise to a new EFT realization. The matching of two theories above and below the
 matching scale is genuinely perturbative. Strong interactions identify with a strongly 
coupled 
theory, which does not decouple. Hence, an EFT realization will arise via a nonperturbative 
matching with the fundamental theory; at the chiral symmetry breaking scale $\Lambda_\chi$, 
quarks leave the ground to pions and kaons, through the nonperturbative matching of ChPT 
with QCD.

\section{ Its descendants}
As symmetries are the foundations of any effective field theory description, we can look 
for extensions of ChPt through its merging with the realization of additional 
symmetries and their breaking. 
A particularly fruitful example is the merging of ChPT with the Heavy Quark Effective 
Theory (HQET) formulated around 1990\cite{HQET}. The additional symmetry is in this case 
the one recovered in the limit of infinitely heavy fermions: the heavy quark spin symmetry. 
The descendants of ChPT are nowadays proliferating, especially after it was realized how   
the interplay of ChPT with lattice QCD can be an invaluable guidance to the theoretical 
interpretation and improvement of lattice calculations; an EFT description can be formulated 
for each purpose, describing the dependence upon the volume, the lattice spacing, the
fermion masses, mimicking and parameterizing the behaviour of a specific lattice 
formulation. With the caveat of a limited energy-range of validity, it offers a 
rigorous theoretical background to interpret physical phenomena on the base of symmetries 
and group theoretical properties.
 
\subsection{The merging with Heavy Quark Effective Theory}

The Dirac theory for spin 1/2 fermions can be reshaped in the limit of an infinitely heavy 
quark. Additional symmetries are restored in this limit, namely the heavy quark spin 
symmetry\cite{HQET}. The merging of HQET with ChPT, gave birth to the effective 
description of baryon interactions, known as Heavy Baryon ChPT (HBChPT)\cite{HBChPT}. 
When baryon number is conserved and taking into account that $m_B\simeq \Lambda_\chi$, 
we can factor out the baryon mass from the total momentum and expand in $1/m_B$. The leading
 order lagrangian describes the interaction of baryons with light meson vector- ($V_\mu$) 
and axial-currents ($A_\mu$) 
\begin{equation}
{\cal L} =  tr\, \bar{B}_v iv\cdot D B_v \,+ \, D\, tr\, \bar{B}_v\gamma^\mu\gamma_5
\{ A_\mu , B_v\}\, 
+ \, F\, tr\, \bar{B}_v\gamma^\mu\gamma_5[ A_\mu , B_v]\, +
\, O\left (\frac{1}{m_B}\right ) + {\cal L}_\pi\, , 
\nonumber
\end{equation}
where 
\begin{equation}
 {B}_v (x) = \frac{1+\displaystyle{\not}v}{2}\,\,B(x)\,e^{im_Bv\cdot x}~~~~~~~~~~F\,+\,D 
\,\,= \,\,g_A\, ,
\end{equation}
with the field ${B}_v (x)$ containing the residual momentum dependence, after the large 
factor $m_Bv$ has been factored out. 
The expansion of HBChPT is therefore a double expansion in 
$1/m_B$ and in $1/\Lambda_\chi$. The first sets the scale of the breaking of heavy quark 
spin symmetry, the second sets the scale of the breaking of chiral symmetry.
An analogous merging gave rise to the description of hadrons with a heavy quark, the EFT 
known as Heavy-Light ChPT (HLChPT)\cite{HLChPT}, which describes the strong interactions of 
D, D* and B, B* mesons with pions.

\subsection{Chiral perturbation theory and lattice QCD}

Field theories can be formulated on a euclidean world-grid, where space 
and time are 
discretized and the unit distance, the lattice spacing, acts as the ultraviolet
 regulator of the theory.
 The euclidean formulation allows for a statistical intepretation of the path integral and 
its treatment with Monte Carlo methods (see ref.\cite{Creutz} for a review). 
Typical lattice simulations of QCD are performed on 
a hypercube with volume $L^3\times L_t$, with spatial extension $L=Na$, temporal 
extension $L_t=N_ta$ and lattice spacing $a$ (in some cases a different lattice spacing 
$a_s\neq a_t$ might be conveniently chosen). For an introduction to lattice field theory and 
lattice QCD see e.g.\cite{LQCDintro}. The lattice formulation allows for a {\it 
first principle} description of a theory, both in the strong-coupling 
(non perturbative) and weak-coupling (perturbative) regimes. 
Ideally ${1}/{L}\ll m_\pi \ll \Lambda_\chi \ll {1}/{a}$ guarantees that pions 
freely move in the lattice box, i.e. their Compton wavelenght is much smaller than $L$, and 
they do not feel the discretization of spacetime. The goal is to be as near as possible to 
the real world limits $L\to\infty$ (the infinite volume limit), $a\to 0$ 
(the continuum limit), and $m_{u,d}\sim m_{u,d}^{phys}$ (the chiral limit for 
$m_{u,d}^{phys}\simeq 0$).  
Typical magnitudes for simulations up to date are $L\sim 2\div 4~fm$, $a\leq  0.1~fm$, 
and $m_{u,d}\leq m_s/2$. Last years have seen an enormous improvement, with simulations at 
lattice spacings down to $a\sim  0.05~fm$ and quark masses as small as $m_{u,d}\sim m_s/8$.

The Symanzik action\cite{Symanzik} was the first example of EFT used to guide a lattice 
calculation, in this case 
to perform the extrapolation to the continuum limit. The generalization of this approach 
is an EFT description that guides the extrapolation to all 
limits, the infinite volume, the continuum and the chiral limit.
During many years the quenched approximation of QCD (QQCD), where the fermionic determinant 
in the path integral is set to a constant, was a forced choice for lattice 
calculations\footnote{ The fermionic 
operator is a large sparse matrix of $spins\times colours\times space\times  time$, that 
renders the exact calculation computationally very expensive.}.
On the way to restore the original content of QCD, one can formulate a partially quenched 
(PQQCD) version of it and the corresponding (partially) quenched ChPT\cite{BG,CP}, where sea 
quarks are distinguished from valence quarks and 
added at will to the theory content. Valence quarks are quenched, while sea quarks are 
dynamical. The QCD point is recovered at $N_{sea}=N_{valence}$ and $m_{sea}=m_{valence}$. 
The symmetry group is the graded extension\cite{BG} of the 
chiral group $SU(N)\times SU(N)$ for $N$ flavours:  
$SU(N|N)\times SU(N|N)$, with $N$ valence and ghost quarks in the quenched case, and
$SU(N+K|N)\times SU(N+K|N)$, for $K$ sea quarks and $N$ valence and ghost quarks, in the 
partially quenched extension.
The construction of (P)QChPT, initiated a stream of results which quickly clarified 
how the approximation affects observables and their volume dependence, using 
symmetry arguments, the non-unitarity of the quenched theory, and group theory 
considerations\cite{GP}.
Further experience in the EFT approach {\em \`a la} Symanzik allowed to guide simulations
towards new regimes of masses and volumes, from the usual $p$-regime to the 
$\varepsilon$-regime when approaching the chiral limit: the original theoretical 
formulation\cite{GLeps} was riproposed\cite{GLWeps} in a lattice context.  
From the $p$-regime, with moderately large volumes and masses, 
$m_\pi L, m_\pi L_t \gg 1$, $2\pi /L \ll\Lambda_\chi \ll 1/a$ and $p/\Lambda_\chi$ small,
one enters the $\varepsilon$-regime while lowering the quark masses, where $m_\pi L, 
m_\pi L_t\sim  \varepsilon \ll 1$. Here the zero modes\cite{Smilga} of the Dirac operator 
must be resummed: $m_q\langle \bar{q}q\rangle L^3L_4\leq O(1)$. 

Nowadays, ChPT formulations match every possible lattice strategy, mainly 
depending on the way fermions are included in the lattice action. The physical prediction 
is unique, but not the way a specific lattice formulation extrapolates to the chiral and 
continuum limits.

\section{ Hot Phenomenology}
Hot topics of today phenomenology are those providing a powerful probe of physics beyond 
the standard model. Restricting to (almost light) hadron phenomenology brings me to 
mention topics as rare kaon decays, which provide tests of the unitarity of the CKM matrix, 
and charm physics. The following short sections are meant to recall a few important 
aspects, while for an extended analysis I refer the reader to the existing literature.

\subsection{Rare decays and CKM unitarity}

The semileptonic decay $Kl_3$ is a gold plated kaon decay. It can provide\cite{Cirigliano} 
a precise test of lepton universality, a determination of the amount of SU(2) breaking,
 through mass ratios, and the amount of SU(3) 
breaking, through the determination of the CKM matrix element $V_{us}$.
The rare semileptonic processes $K_L\to \pi^0\nu\bar{\nu}$ and
$K^+\to \pi^+\nu\bar{\nu}$ are crucial channels to probe new physics contributions. 
For the latter processes, the accurate knowledge of the charm mass is crucial. 
All standard model contributions to these processes are being calculated with
 incresing accuracy 
and with use of ChPT for long distance contributions. For an updated overview, visit 
the Kaon 2007 website\cite{KAON07}.
Finally, the radiative decays $K\to \pi\gamma\gamma , \pi\pi\gamma$ can further 
probe the range of validity of ChPT and long distance dynamics.

\subsection{Charm physics}

The physics of charm is as rich as difficult to decipher. 
The charm is not heavy enough $M_c~{\mbox{is not}}\gg\Lambda_\chi$ to use the heavy quark 
expansion with sufficiently high predictive power, and it is not light enough
$M_c\simeq\Lambda_\chi$ to use the chiral 
expansion. However, it is more relativistic than the bottom quark, 
hence its lattice formulation is affected by smaller discretization errors. We need 
$m_c\ll 1/a$ on the lattice, and it is now easy to get $m_c a\sim 1/2$\cite{Davies}.
What is further needed? Two points are worth to be mentioned: i) a more accurate 
determination of the charm mass (to the percent level), and ii) the prediction of the 
strong interaction phases of D-meson decays which probe CP violation and are indirect 
probes of physics beyond the standard model.

\section{Conclusive thoughts}

During the last two decades we have reshaped our view of 
quantum field theories. Effective field theories are at the foundation of modern 
quantum field theory, and the effective field theory of low energy QCD has significantly 
contributed to this view.
Where is the future of EFTs? They will probably remain for long the bread and butter of 
field theoretical approaches to many phenomena, not only in particle physics, but widely used
in condensed matter physics. There are clear places in particle physics where the 
formulation of an effective field theory description still needs to be fruitfully improved. 
This is the case for the prediction of the electric dipole moments, tiny observables 
measured at very high precision low energy experiments\cite{EDMexp}.
Can one think of a new hybrid EFT formalism to efficiently describe strong interactions in 
charm decays? or the yet unexplored intermediate regime of baryon densities in neutron stars?
One important role of EFT is undeniably the one of uncovering the possible connection
of (super)gravity theories to a four-dimensional universe. 

\end{document}